\documentclass[twocolumn]{aastex62}
\usepackage{amsmath}
\hypersetup{
	colorlinks	= true,
	linkcolor	= red,
	urlcolor	= cyan,
	citecolor	= blue
}
\usepackage{amssymb}
\usepackage{lipsum}
\usepackage{multirow}
\usepackage{lineno}
\usepackage[version=4]{mhchem}
\usepackage{ragged2e}
\usepackage{gensymb}
\usepackage{tikz}
\usepackage{graphics}
\usepackage[figure,figure*]{hypcap}
\usepackage[para,online,flushleft]{threeparttable}
\usepackage{comment}
\usepackage{tablefootnote}
\usepackage{subscript}

\makeatletter
\newcounter{reaction}
\renewcommand\thereaction{R\arabic{reaction}}
\@addtoreset{reaction}{chapter}
\newcommand\reactiontag%
{\refstepcounter{reaction}\tag{\thereaction}}
\newcommand\reaction@[2][]%
{\begin{equation}\ce{#2}%
\ifx\@empty#1\@empty\else\label{#1}\fi%
\reactiontag\end{equation}}
\newcommand\reaction@nonumber[1]%
{\begin{equation*}\ce{#1}\end{equation*}}
\newcommand\reaction%
{\@ifstar{\reaction@nonumber}{\reaction@}}
\makeatother

\graphicspath{{./images/}}

\turnoffediting

\shortauthors{Hu et al.}

\begin{document}

 
	\title{Narrow loophole for \ce{H2}-dominated atmospheres on habitable rocky planets around M dwarfs}
	
	\correspondingauthor{Renyu Hu}
	\email{renyu.hu@jpl.nasa.gov \\ @2023 All rights reserved. \\ Government sponsorship acknowledged.}
	
	\author[0000-0003-2215-8485]{Renyu Hu}
	\affiliation{Jet Propulsion Laboratory, California Institute of Technology, Pasadena, CA 91109, USA}
	\affiliation{Division of Geological and Planetary Sciences, California Institute of Technology, Pasadena, CA 91125, USA}

  	\author[0000-0000-0000-0000]{Fabrice Gaillard}
	\affiliation{Institut des Sciences de la Terre d'Orléans, CNRS/Université d'Orléans/BRGM, 1a rue de la Férollerie, 45071 Orléans cedex 2, France}

 	\author[0000-0002-1426-1186]{Edwin Kite}
	\affiliation{Department of the Geophysical Sciences, University of Chicago, Chicago, IL 60637, USA}
	
	\begin{abstract}
Habitable rocky planets around M dwarfs that have \ce{H2}-dominated atmospheres, if they exist, would permit characterizing habitable exoplanets with detailed spectroscopy using JWST, owing to their extended atmospheres and small stars. However, the \ce{H2}-dominated atmospheres that are consistent with habitable conditions cannot be too massive, and a moderate-size \ce{H2}-dominated atmosphere will lose mass to irradiation-driven atmospheric escape on rocky planets around M dwarfs. We evaluate volcanic outgassing and serpentinization as two potential ways to supply \ce{H2} and form a steady-state \ce{H2}-dominated atmosphere. For rocky planets of $1-7\ M_{\oplus}$ and early, mid, and late M dwarfs, the expected volcanic outgassing rates from a reduced mantle fall short of the escape rates by $>\sim1$ order of magnitude, and a generous upper limit of the serpentinization rate is still less than the escape rate by a factor of a few. Special mechanisms that may sustain the steady-state \ce{H2}-dominated atmosphere include direct interaction between liquid water and mantle, heat-pipe volcanism from a reduced mantle, and hydrodynamic escape slowed down by efficient upper-atmospheric cooling. It is thus unlikely to find moderate-size, \ce{H2}-dominated atmospheres on rocky planets of M dwarfs that would support habitable environments.
	\end{abstract}
	
	
	\section{Introduction} \label{sec:intro}
	
Rocky planets with \ce{H2}-dominated atmospheres would be ideal targets for atmospheric characterization via transmission spectroscopy, because of their large atmospheric scale height that causes large expected spectral features \citep[e.g.,][]{miller2008atmospheric,seager2010exoplanet,greene2016characterizing}. A moderately irradiated rocky planet with a \ce{H2}-dominated atmosphere may have surface pressure and temperature consistent with liquid water \citep{pierrehumbert2011hydrogen,wordsworth2012transient,ramirez2017volcanic,koll2019hot}. Such potentially habitable worlds sustained by \ce{H2}-dominated atmospheres, if they exist around M dwarfs, would unlock the opportunity to study extrasolar habitability with spectroscopy \citep[e.g.,][]{seager2013biosignature}, as TESS and ground-based surveys find temperate rocky planets around M dwarfs \citep[e.g.,][]{sebastian2021speculoos,kunimoto2022predicting}, and JWST provides the sensitivity to analyze any \ce{H2}-dominated atmospheres on them with a wide spectral coverage \citep[e.g.,][]{batalha2018strategies}. Here we ask: Are such worlds likely?

To have surface liquid water, the \ce{H2}-dominated atmosphere cannot be much larger than $\sim10$ bar, because the surface temperature is primarily a function of the size of the atmosphere (this is valid for the stellar irradiation of $200-1400$ W m$^{-2}$ \citep{koll2019hot}). The exact size and irradiation limit depends on the cloud albedo effect \citep[e.g.,][]{popp2015initiation}. This moderate-size atmosphere is much smaller than the massive \ce{H2}-dominated atmospheres proposed to explain the sub-Neptune-sized low-density planet population, which are typically 1\% planet mass or $>10^4$ bar \citep[e.g.,][]{rogers2023conclusive}. A 10-bar \ce{H2} atmosphere would only add $<\sim0.1\ R_{\oplus}$ to the planetary radius, which can be accommodated by typical uncertainties in planetary mass,  radius, and Fe content \citep{luque2022density}. Also, the temperate rocky planets will have a solid surface, as opposed to the sub-Neptunes that may have a permanent magma ocean \citep{kite2020exoplanet}. Because the permeability of the crust decreases dramatically with increasing depth \citep{manning1999permeability}, any post-formation source of \ce{H2} must come from shallow fresh crust via either volcanic outgassing or crustal alteration processes such as serpentinization.

Meanwhile, temperate planets around M dwarfs receive intense high-energy irradiation from host stars because of their close-in orbits, and this intense irradiation can drive hydrodynamic escape from a \ce{H2}-dominated atmosphere \citep[e.g.,][]{salz2016energy,kubyshkina2018grid,kubyshkina2018overcoming}. The high-energy irradiation will be measured by a bevy of new spacecraft \citep{ardila2022star,france2023colorado}.
We are thus motivated to determine the lifetime of a moderate-size \ce{H2} atmosphere -- permitting surface liquid water -- on a large rocky planet orbiting an M dwarf against hydrodynamic escape, and evaluate the geologic processes that could resupply the \ce{H2} atmosphere.

\section{Atmospheric Escape} \label{sec:escape}

The hydrodynamic escape rate, $f_{\rm es}$ (kg s$^{-1}$), can be approximated by the energy-limited escape rate formula,
\begin{equation}
    f_{\rm es} = \frac{\eta_{\rm es}(F_{\rm X}+F_{\rm EUV})\pi R_{\rm p}^3 a^2}{K G M_{\rm p}},
    \label{eq:es}
\end{equation}
where $F_{\rm X}$ and $F_{\rm EUV}$ are the stellar fluxes in X-ray (5-100 \AA) and extreme ultraviolet (EUV, 100-1240 \AA), $a\ge1$ is the ratio between the X-ray/EUV absorbing radius and the (optical) planetary radius, $K\le1$ is a factor that accounts for the Roche lobe effect \citep{erkaev2007roche}, and $\eta_{\rm es}$ is the escape efficiency. Recent hydrodynamic escape models find the escape efficiency to be in a range of $0.1-0.25$ for solar-abundance atmospheres \citep{salz2016energy} and Equation (\ref{eq:es}) is a good approximation of the full hydrodynamic calculations in the Jeans escape parameter regime for temperate rocky planets \citep[Jeans escape parameter $=25-60$,][]{kubyshkina2018grid,kubyshkina2018overcoming}. For a conservative estimate of the escape rate, we adopt $\eta_{\rm es}=0.1$, $a=1$, and $K=1$.

As shown in Table~\ref{tab:escape}, we pick GJ~832, GJ~436, and TRAPPIST-1 as the representative stars for early M, mid M, and late M dwarfs. Their emission spectra in X-ray, Lyman-$\alpha$, FUV, and NUV bands have been measured, and their emission spectra and fluxes in the EUV band can be inferred from these measurements \citep{peacock2019predicting1,peacock2019predicting2}. We find that the lifetime of a 10-bar \ce{H2} atmosphere on a rocky planet that receives Earth-like insolation from these stars would be uniformly $<0.1$ Ga. Thus, a source of \ce{H2} would be needed to maintain such an atmosphere.


\begin{table*}[]
    \centering
    \begin{tabular}{ll|ll|lll|lll|lll}
    \hline\hline
    Star & Type & \multicolumn{2}{l|}{$F$ (erg s$^{-1}$ cm$^{-2}$)} & \multicolumn{3}{l|}{$f_{\rm es}$ ($10^4$ kg s$^{-1}$)} & \multicolumn{3}{l|}{Life of 10-bar atmos (Gyr)} & \multicolumn{3}{l}{Required $x_{\rm H}$ (wt \%)} \\
    & & X-ray & EUV & 1 $M_{\oplus}$ & 3 $M_{\oplus}$ & 7 $M_{\oplus}$ & 1 $M_{\oplus}$ & 3 $M_{\oplus}$ & 7 $M_{\oplus}$ & 1 $M_{\oplus}$ & 3 $M_{\oplus}$ & 7 $M_{\oplus}$ \\ 
    \hline
    GJ 832 & M1.5 & 2.17 & 149 & 3.1 & 2.3 & 2.2 & 0.05 & 0.07 & 0.09 & 9.2 & 1.7 & 0.35 \\
    GJ 436 & M3.5 & 8.71 & 229 & 4.9 & 3.6 & 3.4 & 0.03 & 0.04 & 0.06 & 14 & 2.7 & 0.54 \\
    TRAPPIST-1 & M8 & 171 & 1097$^*$ & 26 & 19 & 18 & 0.006 & 0.008 & 0.01 & 77 & 14 & 2.9 \\
    \hline\hline
    \end{tabular}
    \caption{Escape rates and lifetimes of a 10-bar \ce{H2} atmosphere, and the required hydrogen content in magma for degassing to sustain this atmosphere, on a hypothetical rocky planet of an M dwarf. The distance between the planet and the star results in a bolometric stellar flux the same as Earth's insolation (i.e., the 1-AU equivalent distance). The X-ray fluxes (5-100\AA) are measured by XMM-Newton \citep{loyd2016muscles,ehrenreich2015giant,wheatley2017strong} and the EUV fluxes (100-1240\AA) are based on PHOENIX synthetic spectra guided by FUV and NUV observations \citep{peacock2019predicting1,peacock2019predicting2}. $^*$\cite{bourrier2017reconnaissance} reported a much lower EUV flux (126 erg s$^{-1}$ cm$^{-2}$ at TRAPPIST-1 e) based on Lyman-$\alpha$ measurements, but using this lower value does not change the conclusion of this paper.}
    \label{tab:escape}
\end{table*}

\section{Volcanic Outgassing} \label{sec:outgassing}

We first consider volcanic outgassing as the source of \ce{H2} \citep[e.g.,][]{liggins2020can}. The volcanic outgassing rate, $f_{\rm og}$ (kg s$^{-1}$), can be modeled by the following equation,
\begin{equation}
    f_{\rm og}=\eta_{\rm og} V x_{\rm H},
\end{equation}
where $V$ is the rate of magma generation, $x_{\rm H}$ is the hydrogen content (wt \%) of magma that degasses, and $\eta_{\rm og}$ is the outgassing efficiency. We do not expect the outgassing efficiency to be close to unity because, even though extrusive volcanism (magma that reaches and degases at the planetary surface) can probably degas effectively (but often not completely), intrusive volcanism (magma that does not reach the surface) probably degases poorly, especially for \ce{H2} (to be detailed later in this section). For Earth, the extrusive:intrusive ratio is typically 3:1 to 10:1 \citep{white2006long}, and so as a fiducial value, we assume $\eta_{\rm og}=0.1$.

The rate of volcanism can be estimated for a rocky planet by modeling its thermal evolution history. We adopt the geodynamics model of \cite{kite2009geodynamics} for the rate of volcanism, which used a melting model from pMELTS \citep{ghiorso2002pmelts} for the plate tectonic mode and \cite{katz2003new} for the stagnant lid mode (Table~\ref{tab:volanism}). Focusing first on the planets around field M dwarfs, we take the 4-Gyr age values for the rate of volcanism. The values for the plate tectonic and stagnant lid modes are similar. Detailed models of mantle convection in the stagnant-lid regime predict that volcanism would cease much sooner than what Table~\ref{tab:volanism} indicates \citep{noack2017volcanism,dorn2018outgassing}, but this model uncertainty does not impact the conclusion of this paper. For volcanic outgassing to sustain the atmosphere, it is required that $f_{\rm es}=f_{\rm og}$. With $f_{\rm es}$ and $V$, we derive the required $x_{\rm H}$ and list the values in Table~\ref{tab:escape}.

\begin{table}
    \centering
    \begin{tabular}{l|lll|lll}
    \hline
    \hline
    Mode & \multicolumn{3}{l|}{Plate tectonics} & \multicolumn{3}{l}{Stagnant lid} \\
    Age (Gyr) & 2 & 4 & 6 & 2 & 4 & 6 \\
    \hline
    1 $M_{\oplus}$, 1 $R_{\oplus}$ & 8 & 1.5 & 0.5 & 7 & 1.5 & 0 \\
    3 $M_{\oplus}$, 1.3 $R_{\oplus}$ & - & 2 & 0.7 & - & 2.5 & 0 \\
    7 $M_{\oplus}$, 1.7 $R_{\oplus}$ & - & 4 & 1 & - & 3.5 & 0.7 \\
    \hline
    \hline
    \end{tabular}
    \caption{Rate of volcanism (the mass of magma production divided by the mass of planet, in unit of current Earth's value $3.7516\times10^{-19}$ s$^{-1}$), based on the parameterized model of \cite{kite2009geodynamics}. Dashes correspond to the heat-pipe tectonic regime.}
    \label{tab:volanism}
\end{table}

Arc volcanoes on Earth, which are formed by flux melting caused by release of water from subduction of plates rich in hydrated materials, have magmas that contain $1-7$ wt \% water \citep[e.g.,][]{rasmussen2022magmatic}. The water content in the mid-ocean ridge basalt and the ocean island basalt is lower by $1-2$ orders of magnitude \citep[][]{dixon2002recycled}. Complete outgassing of $1-7$ wt \% water in the form of \ce{H2} would provide an $x_{\rm H}$ of $0.1-0.8$ wt \%. We consider this to be a very generous upper limit; comparing it with Table~\ref{tab:escape} shows that it is very unlikely for volcanic outgassing to sustain the \ce{H2} atmosphere.

\begin{figure}
	\centering
	\includegraphics[width=0.45\textwidth]{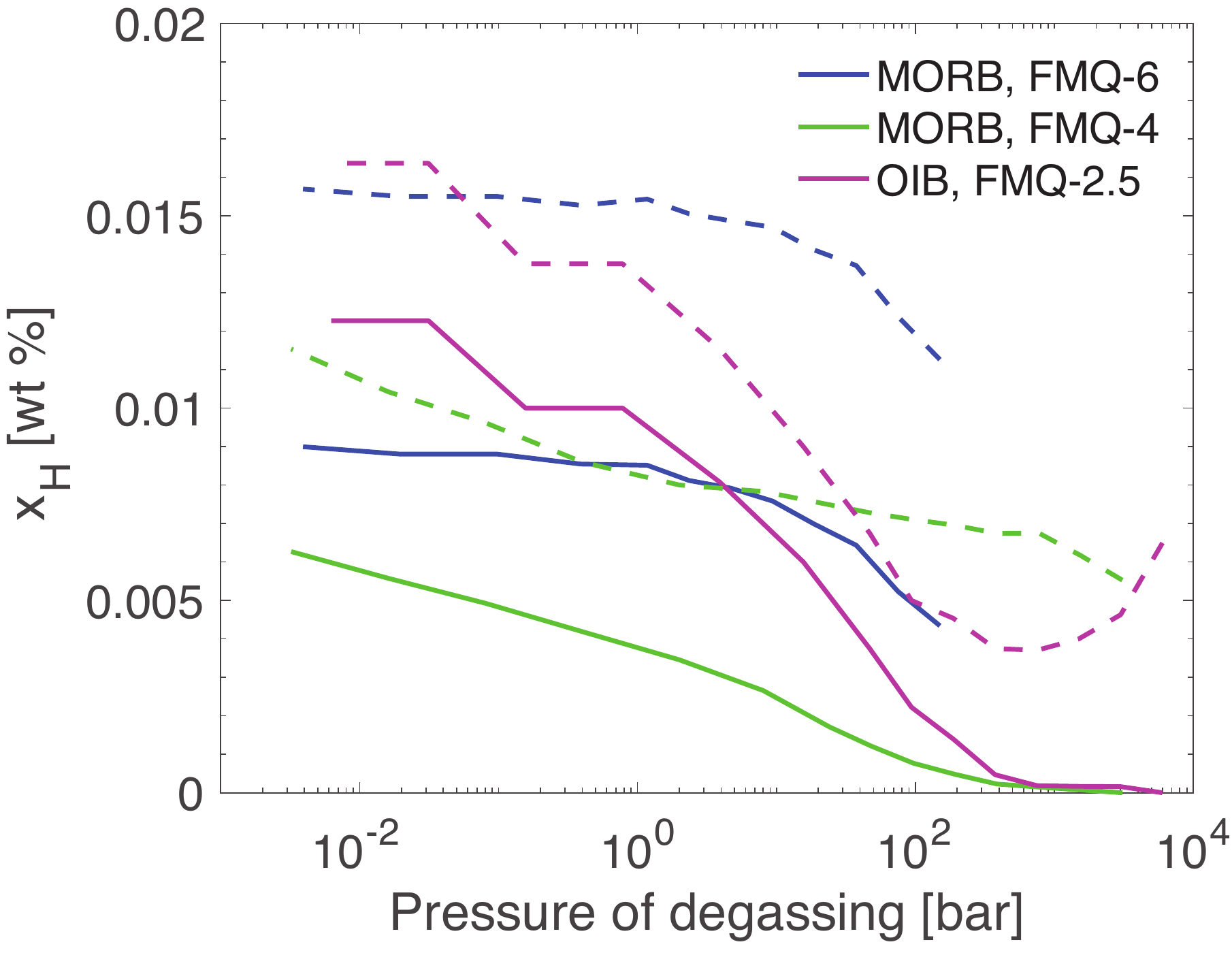}
	\caption{Degassing of \ce{H2} from magma calculated by the magma degassing and speciation model of \cite{gaillard2014theoretical}. We consider mid-ocean ridge basalt (MORB) with bulk \ce{H2O} content of 0.19 wt \% and bulk \ce{CO2} content of 0.16 wt \%, degassing at the oxygen fugacities of FMQ-6 (corresponding to an undifferentiated planet) and FMQ-4 (corresponding to modern Mars), as well as ocean island basalt (OIB) with bulk \ce{H2O} content of 1 wt \% and bulk \ce{CO2} content of 0.3 wt \%, degassing at the oxygen fugacities of FMQ-2.5. The solid lines count the degassing of \ce{H2} and the dashed lines count the degassing of both \ce{H2} and \ce{CO} (with CO expressed in terms of its indirect effect on atmospheric \ce{H2}, see text).}
	\label{fig:degas}
	\end{figure}

The hydrogen content of the magma for degassing depends on the oxygen fugacity of the magma and the pressure at which degassing takes place. We use the magma degassing and speciation model of \cite{gaillard2014theoretical} to calculate $x_{\rm H}$ for the typical volatile content of terrestrial magmas and a wide range of oxygen fugacities (Figure~\ref{fig:degas}). The \ce{H2} content is higher for a more reducing magma and when degassing at a lower pressure. In addition to counting the \ce{H2} degassing, one may also include the potential for atmospheric photochemistry to post-process \ce{CO} to form \ce{H2}, via \ce{CO + H2O -> CO2 + H2}. The complete post-processing means that degassing 1 mole \ce{CO} would be equivalent to degassing 1 mole \ce{H2}, and this situation is shown as dashed lines in Figure~\ref{fig:degas}. However, $x_{\rm H}$ predicted by the geochemical model, even when including the CO conversion, is at least one order of magnitude lower than the minimum required $x_{\rm H}$ for a 7-$M_{\oplus}$ planet around an early M dwarf (Table~\ref{tab:escape}). This again indicates that volcanic outgassing is unlikely to sustain an \ce{H2} atmosphere.

\section{Serpentinization} \label{sec:serp}

We turn to serpentinization as an alternative source of \ce{H2}. Serpentinization is water-rock reactions between warm water and mafic and ultramafic rocks (usually olivine-rich) in the fresh crust. This process probably occurs on all rocky planets with liquid water, and it may have produced \ce{H2}-rich water on early Earth \citep{sleep2004h2} and \ce{H2} and \ce{CH4} on early Mars and on modern Enceladus \citep{oze2005have,chassefiere2011constraining,batalha2015testing,zandanel2021short}. 

For an upper bound of the \ce{H2} production rate from serpentinization, we assume that 1 mole \ce{H2} is produced for every 3 moles \ce{Fe^2+} oxidized, as the process can be written as \ce{H2O + 3FeO -> H2 + Fe3O4}. We also assume that the fresh crust is entirely composed of olivine, \ce{(Mg_{0.9}Fe_{0.1})_2SiO4}, and all \ce{Fe^2+} is used by serpentinization to produce \ce{H2}. The olivine has a molar mass of 146.9 g, and it contains 0.2 moles \ce{Fe^2+}, corresponding to 0.067 moles \ce{H2}, or a mass of 0.13 g. The corresponding $x_{\rm H}$ is thus $0.13/146.9=0.09$ wt.~\%. In reality, the fresh crust may not be entirely composed of olivine, and the rate of serpentinization is limited by the rate of dissolution of olivine in water \citep{oze2007serpentinization}, which is a function of temperature, pH, water/rock ratio, and the Fe/Mg composition of olivine \citep{wogelius1992olivine, allen2003compositional}, as well as by the extent of fracturing of the crust \citep{vance2007hydrothermal}. We thus expect the actual $x_{\rm H}$ provided by serpentinization to be much smaller than 0.09 wt.~\%. However, even this generous upper bound falls short of the required $x_{\rm H}$ by at least a factor of 4 (Table~\ref{tab:escape}). It is thus also unlikely that serpentinization would sustain a moderate-size \ce{H2} atmosphere on rocky planets around M dwarfs. 

	\begin{figure*}
	\centering
	\includegraphics[width=0.8\textwidth]{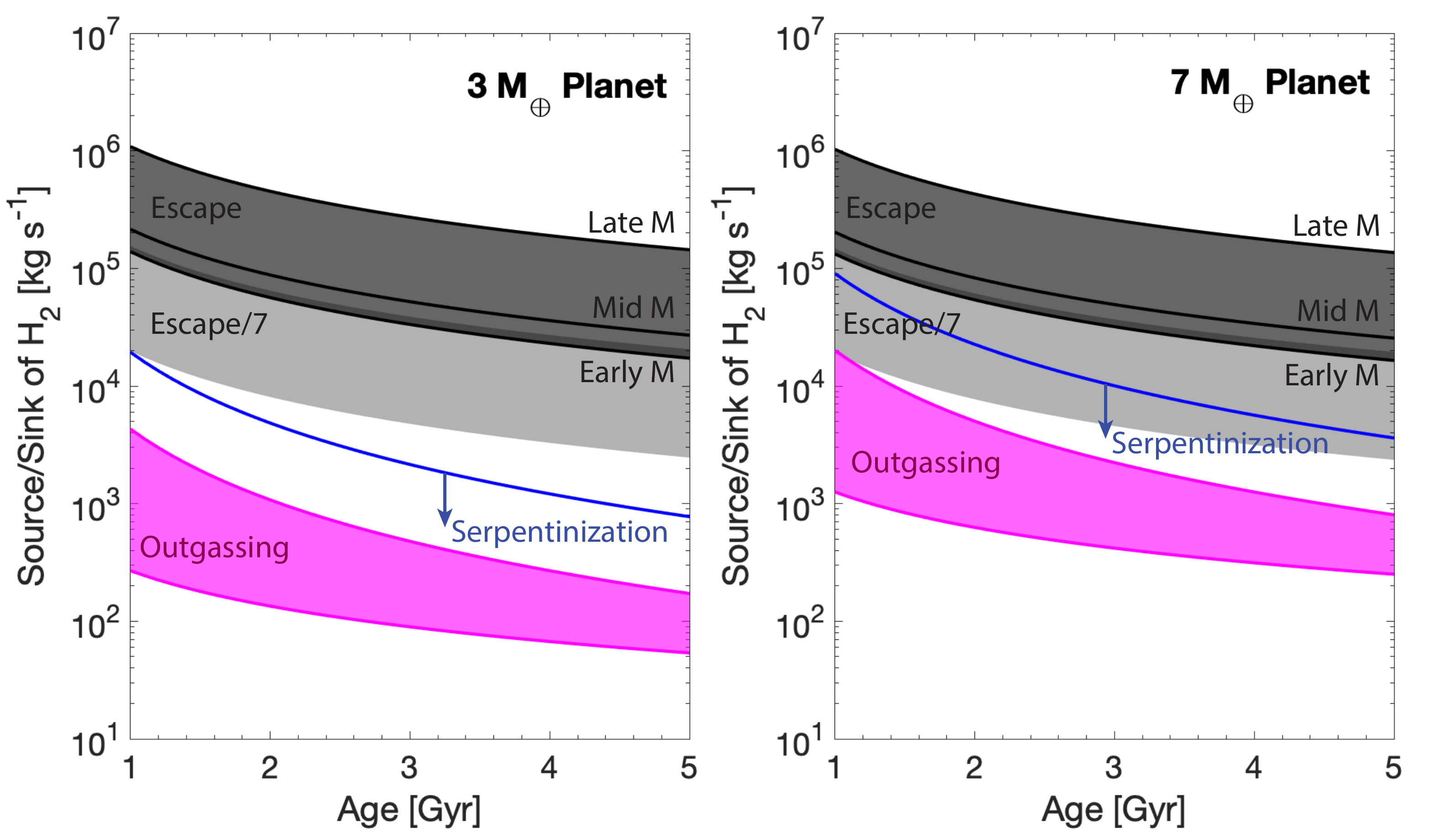}
	\caption{Comparison between sources and sinks of \ce{H2} on rocky planets around M dwarfs. The grey areas show the range of escape rates depending on the type of the host star. The dark grey area is for a planet that locates at the 1-AU equivalent distance, while the light grey area is for a planet that locates 2.6 times farther away (i.e., receiving 7 times less irradiation). The outgassing rates encapsulate the plausible range from a highly reduced mantle (informed by Figure~\ref{fig:degas}), with the lower bound corresponding to $x_{\rm H}=0.005$ wt \% and $t^{-1}$ scaling, and the higher bound corresponding to $x_{\rm H}=0.02$ wt \% and $t^{-2}$ scaling. The rate of serpentinization is a very generous upper limit (Section \ref{sec:serp}).}
	\label{fig:compare}
	\end{figure*}

\section{Age and Distance Dependency} \label{sec:age}

So far we have assumed 4 Gyr for the planet age, which broadly corresponds to the field M dwarfs. Now we consider the age dependency of the sources and sinks of \ce{H2} and see if a steady-state \ce{H2} atmosphere would be possible on younger planets. \cite{richey2019hazmat} recently presented the NUV, FUV, and X-ray fluxes of M dwarfs in their habitable zones as a function of age, and meanwhile, the EUV fluxes follow a similar age dependency as their FUV fluxes \citep{peacock2020hazmat}. We adopt an age dependency of $t^{-0.9}$ for the X-ray fluxes and $t^{-1.3}$ for the EUV fluxes. Meanwhile, the rate of volcanism can be much higher for young planets, when the heat flux from the planetary interior is higher. We explore an age dependency that varies between $t^{-1}$ \citep[based on the model for Earth in][]{schubert2001mantle} and $t^{-2}$ \citep[based on the model for large rocky planets of][]{kite2009geodynamics}. We consider an age as young as 1 Gyr. Before that, the planet could have a residual primordial \ce{H2} atmosphere \citep{kite2020exoplanet} and the stellar high-energy output may have different age dependencies \citep{richey2019hazmat}. As shown in Figure~\ref{fig:compare}, it remains unlikely for volcanic outgassing or serpentinization to compensate for the intense atmospheric escape of \ce{H2} experienced by rocky planets of M dwarfs from 1 to 5 Gyr. 

How about a planet that is located further away from the star than the 1-AU equivalent distance? Moving the planets 2.6 times further away (or receiving 7 times less irradiation, or $\sim200$ W m$^{-2}$) would still produce a potentially habitable climate \citep{koll2019hot}, and this would reduce the escape rate by a factor of 7. In this case, the escape rate is comparable to the upper limit of serpentinization (Figure~\ref{fig:compare}). However, the upper limit assumes complete oxidization of \ce{Fe^2+} in the fresh crust and is thus unlikely. The utility of these distant habitable worlds for observations is probably limited, as they are less likely to transit and harder to find than the closer-in planets.

\section{Potential Alternative Mechanisms}
\label{sec:way}

The estimates above show that it is unlikely to sustain moderate-size \ce{H2}-dominated atmospheres on rocky planets around M dwarfs through volcanic outgassing or serpentinization. Here we explore alternative mechanisms that could result in large source fluxes of \ce{H2}.

First, the rate of hydrogen generation during serpentinization is controlled by the Fe content of olivine \citep{klein2013compositional}. In Section~\ref{sec:serp}, we have assumed a Fe:Mg ratio of 1:9, corresponding to the terrestrial value. On Mars, however, the Fe:Mg ratio of crustal olivine can be $\sim$1:1 \citep{koeppen2008global,morrison2018crystal}. Such Fe-rich olivine could result in higher fluxes of \ce{H2} from serpentinization than our estimates by a factor of $\sim5$, making it more likely for serpentinization to meet the \ce{H2} escape flux.

Second, on a planet with plate tectonics but amagmatic spreading, water-rock interaction near the ridge axis could produce \ce{H2}. Our discussion of serpentinization so far assumes that water interacts with the products of volcanism/partial melting. However, water could interact directly with the mantle if it is exposed by amagmatic spreading. This mechanism occurs on Earth today at Gakkel Ridge and Southwest Indian Ridge and has the potential to generate more \ce{H2} because the Fe content of mantle rock is greater than that of the crustal basalt. This mechanism would decouple serpentinization from volcanism and allow serpentinization to continue even after volcanism has shut down. The upper limit of \ce{H2} production from this mechanism is then given by assuming (unrealistically) a 100\% fayalite (\ce{Fe2SiO4}), which would give an equivalent $x_{\rm H}$ of 0.6 wt. \%. Suppose fractures, and therefore hot-water alteration, penetrate as far down into the subsurface on this amagmatic planet as the base of the oceanic crust on our planet, which is probably unrealistic because fractures should self-seal at shallower depths \citep{vance2007hydrothermal}. Then the present-day terrestrial production of 20 km$^3$/year of MORB, with full serpentinization, would correspond to $1\times10^4$ kg s$^{-1}$ of \ce{H2} output. This is on the same order of magnitude as the lower end of the escape rate (Table~\ref{tab:escape}), and could be higher for younger or larger rocky planets.

Third, Earth's heat flux of $\sim0.1$ W m$^{-2}$ is $\sim10\%$ in the form of advective cooling (magma moves upward and cools) and $\sim90\%$ conductive cooling. This implies that the rates of volcanism could be $10\times$ higher without excessively cooling Earth's mantle. Indeed, it has been hypothesized that heat-pipe tectonics occurred early in Earth's history \citep[e.g.,][]{moore2013heat}. Small variations in exoplanet mantle composition or exoplanet mantle volatile content, among other factors, could make melting easier at a given mantle temperature \citep{spaargaren2020influence}, perhaps enabling heat-pipe mode of planetary cooling even for planets that are as old as the Earth. If heat-pipe volcanism occurs then the amount of outgassing and serpentinization could be $\sim10\times$ greater than for an Earth-scaled plate-tectonics model, because $\sim10\times$ more eruptions  would occur.

Fourth, N-, O-, and C-bearing molecules mixed in the \ce{H2}-dominated atmosphere may substantially reduce the escape rate. The escape rate and efficiency calculations have been based on solar-abundance atmospheres \citep{kubyshkina2018grid,kubyshkina2018overcoming}. However, \ce{H2}-dominated atmospheres sustained by volcanism should also have \ce{H2O}, \ce{CO}/\ce{CH4}, and \ce{N2}/\ce{NH3} at the levels that exceed the solar abundance \citep{liggins2022growth}. Recently, \cite{nakayama2022survival} show that with an \ce{N2}-\ce{O2} atmosphere, cooling from atomic line emissions (of \ce{N}, \ce{N+}, \ce{O}, \ce{O+}) and radiative recombination can prevent rapid hydrodynamic escapes for XUV irradiation fluxes that are up to $1000\times$ the modern Earth value. It is thus conceivable that an \ce{H2}-dominated atmosphere richer in N, O, and C would be more stable. The challenge is that the N-, O-, and C-bearing molecules are separated from \ce{H2} by diffusion (typically at $\sim1$ Pa) and can be largely depleted in the upper atmosphere. It is thus unclear whether 10\% non-\ce{H2} species (which would still allow for a low molecular weight atmosphere for transmission spectroscopy) would slow down the hydrodynamic escape sufficiently to achieve long-term stability.

Lastly, there could be transient episodes of high volcanism and serpentinization that support \ce{H2}-dominated atmospheres. A leading hypothesis for why Early Mars sometimes had lakes is that a lot of \ce{H2} was emitted transiently from the subsurface by volcanism or serpentinization \citep{wordsworth2021coupled}. The amount of \ce{H2} needed to warm up Mars by \ce{H2}-\ce{CO2} collision-induced absorption is now well understood \citep{turbet2020measurements}, and the \ce{H2} flux needed is approximately $10^4$ kg s$^{-1}$. A large rocky planet could have $10\times$ the surface area of Mars and thus plausibly $10\times$ the amount of serpentinization. This process-agnostic (but model-dependent) scaling hints at short-term source fluxes sufficient for \ce{H2}-dominated atmospheres on a 7-$M_{\oplus}$ planet (Figure~\ref{fig:compare}).

\section{Summary} \label{sec:summary}

From the analyses presented above, we conclude that rocky planets around M dwarfs rarely have potentially habitable conditions accompanied by \ce{H2}-dominated atmospheres. This is because, forming a potentially habitable environment will require a moderate-size ($\sim10$ bar) atmosphere, but such an atmosphere is removed quickly by stellar X-ray and EUV irradiation, and could only exist on the planet as a steady-state atmosphere with replenishment. However, neither volcanic outgassing nor serpentinization provides the required \ce{H2} source that would maintain such a steady-state atmosphere. Small planets around M dwarfs could have massive \ce{H2} atmospheres, but to have a stable and moderate-size \ce{H2} atmosphere consistent with habitability would require special circumstances such as direct interaction between liquid water and mantle (e.g., near a ridge axis undergoing amagmatic spreading), heat-pipe volcanism from a highly reduced mantle, or hydrodynamic escape quenched by efficient atomic line cooling. These special mechanisms to sustain the moderate-size \ce{H2} atmosphere are generally more effective on large rocky planets (e.g., $\sim7$-$M_{\oplus}$ planets exemplified by LHS~1140~b) than on Earth-sized planets.

The finding here is consistent with the non-detection to date of clear \ce{H2}-dominated atmospheres on rocky planets of M dwarfs via transmission spectroscopy \citep[e.g.,][]{de2018atmospheric,lustig2023jwst}, although these results could also be interpreted as widely occurring photochemical haze that mutes transmission spectral features of \ce{H2}-dominated atmospheres. \cite{swain2021detection} suggested an \ce{H2}-dominated atmosphere on the rocky planet GJ~1132~b based on HST data, but independent data analyses could not confirm their result \citep{mugnai2021ares,libby2022featureless}. The ongoing HST and JWST transmission spectroscopy of small exoplanets of M dwarfs will further test our findings and, potentially, discover exceptional cases.
Meanwhile, \ce{N2}-\ce{CO2} or other high-mean-molecular-weight atmospheres should probably be considered as the default assumption when planning for future spectroscopic observations of rocky planets around M dwarfs. This would require planning more repeated visits of preferred targets for transmission spectroscopy \citep[e.g.,][]{batalha2018strategies}, or turning to thermal emission spectroscopy and phase-curve mapping \citep[e.g.,][]{angelo2017case,kreidberg2019absence,mansfield2019identifying,koll2019identifying,whittaker2022detectability}.
	
\section*{Acknowledgments}
We thank Evgenya Shkolnik and Tyler Richey-Yowell for helpful discussion of stellar evolution. This work was supported in part by the NASA Exoplanets Research Program grant \#80NM0018F0612. E.S.K. acknowledges support from a Scialog grant, Heising-Simons Foundation 2021-3119. Part of this research was carried out at the Jet Propulsion Laboratory, California Institute of Technology, under a contract with the National Aeronautics and Space Administration. 



\end{document}